\documentclass[aps,prl,reprint,preprintnumbers,nofootinbib,superscriptaddress]{revtex4-1}
\pdfoutput=1
\usepackage{natbib}
\usepackage{amsmath}
\usepackage{amssymb}
\usepackage{amsfonts}
\usepackage{hyperref}
\usepackage{epsfig}
\usepackage{graphicx}
\usepackage{slashed} 
\usepackage{mathrsfs} 
\usepackage{subcaption}
\allowdisplaybreaks
\usepackage{url}

\usepackage{color}
\definecolor{myred}{rgb}{0.6,0,0} 
\definecolor{myblue}{rgb}{0,0.2,0.4}
\definecolor{mygreen}{rgb}{0,0.9,0.1}
\definecolor{hc}{rgb}{.9,0.1,0.7}
\definecolor{hcout}{rgb}{.9,0.7,0.9}
\definecolor{Orange}{rgb}{1.,0.65,0.}

\def\beq{\begin{equation}}
\def\eeq{\end{equation}}
\def\bea{\begin{eqnarray}}
\def\eea{\end{eqnarray}}
\def\ba{\begin{array}}
\def\ea{\end{array}}

\def\t1{\tilde{t}_1}
\def\t2{\tilde{t}_2}
\def\b1{\tilde{b}_1}

\newcommand{\bd}{\begin{displaymath}}
\newcommand{\ed}{\end{displaymath}}
\newcommand{\be}{\begin{equation}}
\newcommand{\ee}{\end{equation}}

\def\b{\beta}

\def\q2 {q^2}

\def\t {\times }

\def\bt{\begin{table}}
\def\et{\end{table}}

\catcode`@=11 
\def \gsim{\mathrel{\mathpalette\@versim>}}
\def \lsim{\mathrel{\mathpalette\@versim<}}
\def \@versim#1#2{\lower0.4ex\vbox{\baselineskip\z@skip\lineskip\z@skip
     \lineskiplimit\z@\ialign{$\m@th#1\hfil##\hfil$%
     \crcr#2\crcr\sim\crcr}}}
\catcode`@=12 

\begin{document}
\preprint {HRI-RECAPP-2018-009}

\title{Can SKA--Phase 1 go much beyond the LHC in supersymmetry search?}

\author{Arpan Kar}
\email{E-mail: arpankar@hri.res.in}
\affiliation{Regional Centre for Accelerator-based Particle
  Physics, Harish-Chandra Research Institute, HBNI, Chhatnag Road,
  Jhunsi, Allahabad - 211 019, India}
\author{Sourav Mitra}
\affiliation{Surendranath
  College, 24/2 M. G. ROAD, Kolkata, West Bengal 700009}
\author{Biswarup Mukhopadhyaya}
\affiliation{Regional Centre for Accelerator-based Particle
  Physics, Harish-Chandra Research Institute, HBNI, Chhatnag Road,
  Jhunsi, Allahabad - 211 019, India}
\author{Tirthankar Roy Choudhury}
\affiliation{National Centre for Radio Astrophysics, TIFR, Post Bag 3, Ganeshkhind, Pune 411007, India}

\begin{abstract}
We study the potential of the Square Kilometre Array in the first phase
(SKA1) in detecting
dark matter annihilation signals from dwarf spheroidals in the form of diffuse radio synchrotron.
Taking the
minimal supersymmetric standard model as illustration, we show that it is
possible
to detect such signals for dark matter masses about an order of magnitude
beyond
the reach of the Large Hadron Collider, with about 100 hours of observation
with the SKA1.
\end{abstract}

\maketitle {\em Introduction: The upcoming low-frequency
  radio telescope, Square Kilometre Array Phase~1 (SKA1), can well
  surpass the Large Hadron Collider (LHC) reach in unveiling new
  physics responsible for dark matter (DM). This is shown for the minimal supersymmetric
  standard model (MSSM) where the lightest neutralino ($\chi^0_1$) is
  the DM candidate. While the LHC is unlikely to see
  signatures of supersymmetry (SUSY) for ($m_{\chi^0_1} \gsim$)1 TeV,
  especially for coloured superparticle masses above $\simeq$3
  TeV \cite{LHC_reach, LHC_reach2}, DM annihilation in dwarf spheroidal galaxies (dSph) can lead to radio
  synchrotron emission which clearly rises above the SKA1 detection
  threshold with 10-100 hours of observation, for $m_{\chi^0_1}$ up to
  at least 10 TeV. We scan the MSSM parameter space and
  predict the synchrotron radiation spectra for three such galaxies. 
  for DM annihilation corresponding to the aforesaid SUSY breaking
  scales. Even with conservative parameter values, the SKA1 should
  see signals, for  DM masses one order higher than the reach of the LHC
  with $\int {\cal L} dt = 3000 \mbox{fb}^{-1}$ \cite{LHC_reach, LHC_reach2}.}

The SKA1 is expected to address many important questions in
astrophysics and cosmology \cite{Braun:2015zta}.  It has relevance in
the physics of elementary particles, too.  Foremost in this context is
the issue of DM, provided it is constituted of elementary particle(s).
While there is no unique candidate theory yet, the MSSM shows a
logically satisfactory way to obtain a stable neutral particle,
especially the lightest neutralino ($\chi^0_1$), which satisfies the
requirements of DM. The LHC, however, has not found any signals of it
so far, up to coloured new particle masses $\gsim$ 2 TeV \cite{Sirunyan:2017cwe_Aaboud:2017bac}. On the other
hand, spectra in the multi-TeV range can be phenomenologically
allowed and satisfy all requirements of DM, if one defers judgements
on the somewhat fuzzy issue of naturalness. While the
LHC cannot see such heavy superparticles, and the fate of any future collider
is uncertain,  we show below that the
SKA1 in its first phase itself can detect diffuse radio synchrotron
signals of DM annihilation for such high $m_{\chi^0_1}$.

In dSphs and  galactic clusters, DM-pairs annihilate into standard model (SM) particles
such as $b\bar{b}, t\bar{t}, W^{+}W^{-}, \tau^{+}\tau^{-}....$. The
subsequent cascades produce copious $e^{+}e^{-}$
pairs whose energy distribution is determined by the source
function
\begin{equation}
Q_e(E,r) = \frac{\langle \sigma v \rangle}{2 m_{\chi}^2} \rho^2_{\chi}(r) \sum_{f} {\frac{dN^e_f(E)}{dE}} B_f 
\end{equation}
where $m_{\chi}$ is the DM mass ($m_{\chi^0_1}$ in MSSM),
$\langle \sigma v \rangle$ and $\rho_{\chi}$ are, respectively, the DM
annihilation rate in any aforementioned channel with fraction $B_f$ and
DM density inside the galaxy. Here we have used the NFW profile for Draco
\cite{Colafrancesco:2006he, McDaniel:2017ppt} and Ursa Major~II
\cite{Natarajan:2013dsa}, and the Einasto profile for Segue~1
\cite{Natarajan:2015hma}, with the parameters of the
  profile chosen such that they are consistent with the kinematical
  observations of these galaxies. $d N^e_f(E) / dE$ is the energy
distribution of the $e^{\pm}$ per annihilation.

Prediction of the synchrotron signal produced by these $e^{+}e^{-}$
pairs requires tracking their propagation through galactic media.
Accounting for the energy loss via various electromagnetic
processes as well as diffusion, the steady state distribution $dn/dE(E, r)$
of the $e^{\pm}$ as a function of energy $E$ and distance $r$ from the
centre of the galaxy can be obtained by solving the equation
\cite{McDaniel:2017ppt, Colafrancesco:2005ji}
\begin{equation}
D(E) \nabla^2 \left(\frac{dn}{dE}\right) +
\frac{\partial}{\partial E}\left(b(E) \frac{dn}{dE}\right) +
Q_e(E,r) = 0
\end{equation}
where $b(E)$ denotes the energy loss due to various
radiative processes like the inverse Compton scattering, synchrotron
radiation, Coulomb losses and bremsstrahlung. $D(E)$ is the diffusion
parameter which is assumed to have the Kolmogorov form $D = D_0
E^{\gamma}$ \cite{Colafrancesco:2006he, McDaniel:2017ppt, Natarajan:2013dsa, Natarajan:2015hma}, $D_0$ being the diffusion coefficient.  One finally
obtains the frequency spectrum of observed photons by folding $dn/dE$ 
with the synchrotron power spectrum, for which a
simplified expression is available for frequencies above the
gyro-frequency and plasma frequency \cite{McDaniel:2017ppt,
  Colafrancesco:2005ji, Beck:2015rna}.

Nearby ultra-faint dSphs are appropriate for studying such diffuse
radio signals, as opposed to usual galaxies and clusters, as their low
star formation rate minimises the uncertain contribution of
astrophysical processes. Their relative proximity (most of them are
satellites of the Milky Way) and high DM content, as inferred from the
observed mass-to-light ratios within their half-radii, are of further
advantage.  Some of these dSphs have been observed using existing
radio telescopes with the aim of recording such diffuse emission,
although no signal has materialsed so far \cite{Natarajan:2013dsa,
  Natarajan:2015hma, Spekkens:2013ik, Regis:2017oet_Regis:2014tga}.  The ultra-faint nature of
these galaxies necessitates a more sensitive telescope like
SKA1 for detecting the radio synchrotron signal
\cite{Colafrancesco:2015ola}. Here we predict the diffuse
signal considering the parameters for Draco dSph (mainly because the
various relevant parameters like the $J$-factor are better constrained
for this object \cite{Geringer-Sameth:2014yza}), though even higher
flux is expected out of the nearer dSphs such a Segue~1 and Ursa
Major~II, as shown later in this note.

For a $\chi^0_1$ DM in MSSM, the observed radio flux (obtained via the
velocity averaged quantity $\langle \sigma v \rangle$ (calculated using micrOMEGAs 4.3.1 \cite{Belanger:2001fz})) depends
on not only $m_{\chi^0_1}$ but also the particle spectrum and other
MSSM parameters that determine the annihilation rates and branching
ratios, and also the energy of $e^{\pm}$ transported across the dSph. Some recent works \cite{McDaniel:2017ppt, Colafrancesco:2015ola, Beck:2017hkp} have treated
$m_{\chi}$ and $\langle \sigma v \rangle$ as two free parameters,
and studied the consequences of different `dominant' annihilation
channels. We instead select various MSSM benchmark regions, especially
those with heavy superparticles undetectable at the LHC \cite{LHC_reach, LHC_reach2}, and use the
full dynamics of the model in terms of the emergent annihilation
channels. These benchmarks are listed in Tables
\ref{Table_benchmark_1} and \ref{Table_benchmark_2}. There are four
broad classes.  (A) has all squarks/gluinos and sleptons well above
LHC detection limits, but with a hierarchy between squarks and
sleptons. (B) includes somewhat lighter but still undetectable
superparticles, but with no hierarchy between coloured and colourless
ones. (C) and (D) have similar spectra as in (A) and (B) but with
lighter top squarks in each case. (E) and (F) correspond to ultra-high
$\chi^0_1$ masses close to 10 TeV.  These regions identify DM candidates beyond the
commonly conceived domain of naturalness. Further categories within
each class reflect different combinations of other MSSM parameters
which drive annihilation in different channels. In addition, spectra
with $\chi^0_1$ beyond the LHC detection limit have been juxtaposed
with relatively light ones for comparison. All benchmarks satisfy the
constraints coming from collider \cite{Sirunyan:2017cwe_Aaboud:2017bac, Aaboud:2018jiw_Sirunyan:2018ubx, Aaboud:2017sjh} and direct DM searches \cite{Aprile:2017iyp_PhysRevLett.118.251302}, relic
density \footnote{We have demanded that the relic density should lie within stipulated {\it upper and lower limits}, as is expected in a single-component DM scenario. The $\langle \sigma v \rangle$ required for relic density calculation needs to be evaluated in a way appropriate for the decoupling temperature, including co-annihilation channels.} \cite{Ade:2013zuv_Harz:2016dql_Klasen:2016qyz}, lightest neutral Higgs mass \cite{Allanach:2004rh} (calculated at the two-loop
level), flavour physics \cite{Aaij:2017vad_Sandilya:2017mkb}, $(g-2)_{\mu}$ \cite{muon_g_2} etc.

\begin{table}[ht!]
\centering
\begin{tabular}{|c|c|c|c|c|c|c|c|}
\hline 
Cases & $m_{\widetilde{l}}$ & $m_{\widetilde{Q}_{1,2,3}}$ & $m_{\widetilde{q_R}}$
& $A_{t}$ & $m_{\widetilde{t_R}}$ & $m_{\widetilde{g}}$ \\ 
 & (TeV) & (TeV) & (TeV) & (TeV) & (TeV) & (TeV) \\
\hline
A & 5-5.5  & 10 & 10 & -2 & 10 & 10\\ 
\hline
B & 4-4.5  & 5 & 5 & -4 & 5 & 5\\
\hline
C & 4-4.5  & 5 & 5 & -4 & 1.95 & 5\\
\hline
D & 5-5.5  & 10 & 10 & -2 & 2.05 & 10\\
\hline
E & 10  & 10 & 10 & -2 & 10 & 10\\
\hline
F & 15  & 20 & 20 & -1 & 20 & 20\\
\hline
\end{tabular}
\caption{Parameters characterizing different classes of benchmarks. 
  $m_{\widetilde{l}}$ stands for  all three slepton families, except
  in A1c (see Table \ref{Table_benchmark_2}) where
  $m_{\widetilde{\tau_1}}$ has been
  fixed at = 1.03 $m_{\chi^0_1}$ to emphasize the $\tau^+ \tau^-$
  annihilation channel.}
\label{Table_benchmark_1}
\end{table}

\begin{table*}[ht!]
\begin{center}
\def\arraystretch{1.5}%
\begin{tabular}{|c|c|c|c|c|c|c|c|}
\hline 
Model & $M_1$(GeV) & $M_2$(GeV) & $\mu$(GeV) & $M_A$(GeV) & $tan\beta$ & annihilation channel & $m_{\chi^0_1}$(GeV) \\ 
\hline
A1a & 1020.0 & 2000.0  & 1180.0 & 2113.0 & 20 & $b\bar{b}(85\%)$, $\tau^+\tau^-(14\%)$ &  1000.6\\ 
\hline
A1b & 2097.2 & -3536.3  & 1134.8 & 7022.6 & 20 & $W^+W^-(55\%)$, $ZZ(45\%)$ & 1163.0\\ 
\hline 
A1c & 1030.0 & 3000.0  & 1150.0 & 2200.0 & 5 & $\tau^+\tau^-(38\%)$, $t\bar{t}(37\%)$, $b\bar{b}(22\%)$, $W^+W^-(1.7\%)$, $ZZ(1.3\%)$ & 1006.7\\
\hline 
A2a & 3932.4 & 3645.7  & -3427.5 & 7001.2 & 20 & $b\bar{b}(76\%)$, $\tau^+\tau^-(15\%)$, $t\bar{t}(4\%)$, $W^+W^-(3\%)$, $ZZ(2\%)$ & 3459.4\\  
\hline 
A2b & 5537.0 & -2976.8  & -3372.3 & 6517.9 & 20 & $W^+W^-(91\%)$, $b\bar{b}(7.6\%)$, $\tau^+\tau^-(1.4\%)$ & 3085.4\\ 
\hline
A2c & 4477.6 & 3977.5  & 4330.9 & 8293.9 & 20 & $b\bar{b}(53.4\%)$, $W^+W^-(35\%)$, $\tau^+\tau^-(11\%)$ & 4090.6\\ 
\hline
A3 & -312.0 & 1000.0  & 400.0 & 690.8 & 10 & $t\bar{t}(79.4\%)$, $b\bar{b}(16.3\%)$, $\tau^+\tau^-(2.2\%)$, $W^+W^-(1\%)$ & 302.0\\ 
\hline
B1a & -1013.8 & 2022.0  & 1150.0 & 2113.0 & 20 & $b\bar{b}(72\%)$, $t\bar{t}(16\%)$, $\tau^+\tau^-(12\%)$ & 1000.0\\
\hline
B1b & -3884.7 & 3550.0  & 1132.7 & 3627.7 & 20 & $W^+W^-(55\%)$, $ZZ(45\%)$ & 1153.4\\ 
\hline
B2a & -3485.5 & 4177.9  & 3354.3 & 6820.0 & 20 & $b\bar{b}(76\%)$, $\tau^+\tau^-(15\%)$, $W^+W^-(3\%)$, $t\bar{t}(3\%)$, $ZZ(2.8\%)$ & 3368.0\\   
\hline 
B2b & -3930.2 & -2598.1  & -2957. & 5752.4 & 20 & $W^+W^-(94.2\%)$, $b\bar{b}(5\%)$ & 2662.0\\  
\hline
B3 & -295.0 & 1000.0  & 400.0 & 668.0 & 20 & $b\bar{b}(50\%)$, $t\bar{t}(42\%)$, $\tau^+\tau^-(7\%)$ & 286.0\\  
\hline
C & -1012.0 & 3000.0  & 2000.0 & 2033.5 & 10 & $b\bar{b}(63.6\%)$, $t\bar{t}(26\%)$, $\tau^+\tau^-(10.2\%)$ & 1012.4\\
\hline
D & 1015.0 & 3000.0  & 2000.0 & 2047.0 & 10 & $b\bar{b}(60\%)$, $t\bar{t}(30\%)$, $\tau^+\tau^-(10\%)$ & 1015.4\\
\hline
E & 8600.0 & 10000.0  & 8500.0 & 17035.0 & 20 & $b\bar{b}(79.1\%)$, $\tau^+\tau^-(18.3\%)$, $t\bar{t}(2.5\%)$ & 8498.0\\
\hline
F & 11000.0 & 9700.0  & 9965.0 & 20000.0 & 20 & $b\bar{b}(78.5\%)$, $\tau^+\tau^-(17.8\%)$, $t\bar{t}(1.9\%)$, $W^+W^-(1.6\%)$ & 9947.4\\
\hline
\end{tabular}
\caption{Parameters in different benchmark Models within the classes listed in 
Table \ref{Table_benchmark_1}, and the corresponding DM masses and annihilation channels.}
\label{Table_benchmark_2}
\end{center}
\end{table*}

\begin{figure}
\includegraphics[height=0.35\textwidth, angle=0]{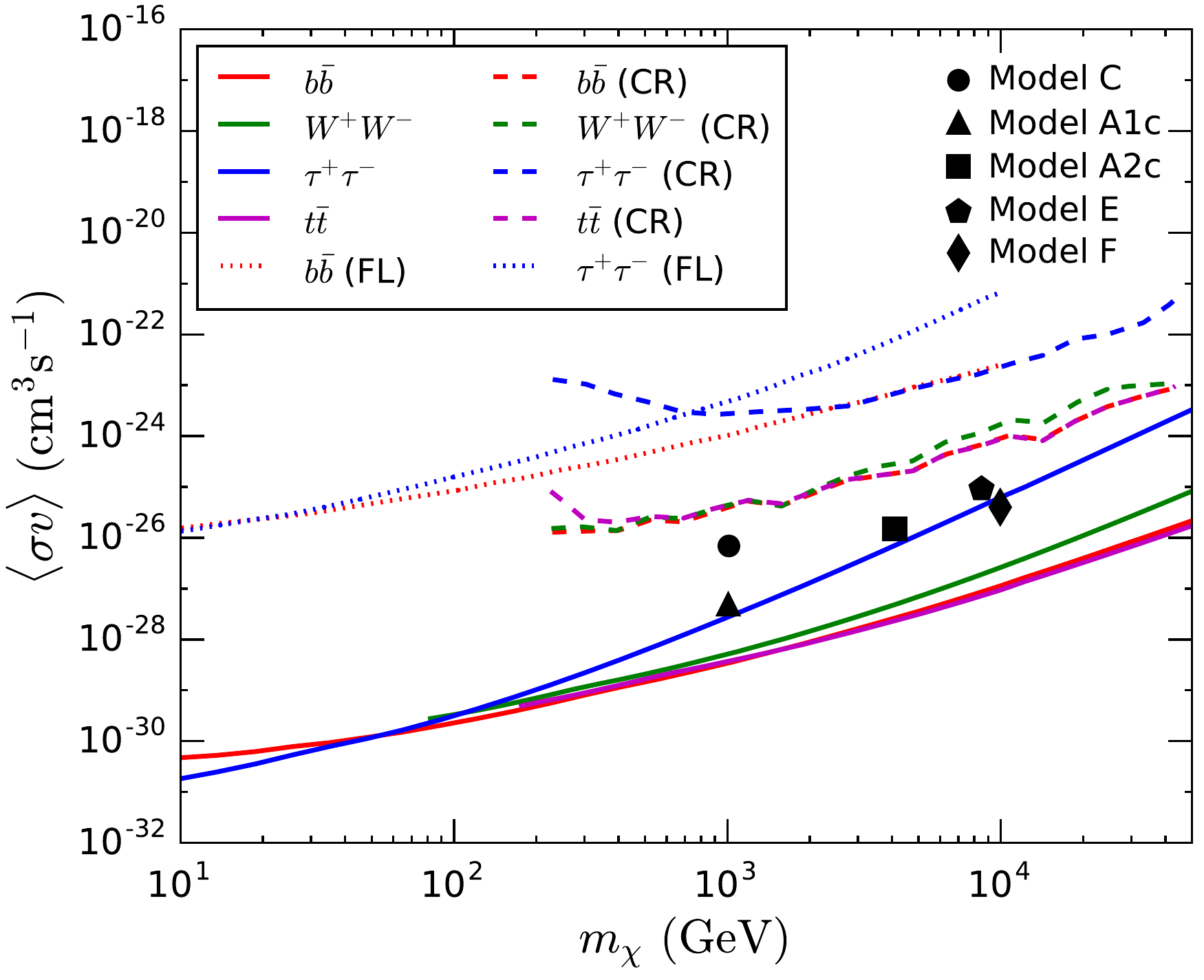}
\caption{Lower limits (solid lines) of observability of radio flux
  from Draco in the $\langle \sigma v \rangle - m_{\chi}$
  ($m_{\chi^0_1}$ in MSSM) plane at SKA1 with 100 hours, for various
  DM annihilation channels.  Dashed and dotted lines denote the
  corresponding 95\% C.L. upper limits from cosmic-ray (CR) antiproton
  observation and 6 years of Fermi LAT (FL) data \cite{fermiLAT} respectively.}
\label{sv_mchi}
\end{figure}

Figure \ref{sv_mchi} shows the minimum $\langle \sigma v \rangle$
required in various channels for detection with 100 hours (bandwidth = 300 MHz) at the SKA1,
for the dSph Draco. The corresponding annihilation channel has to
dominate in each case, for the lower limit to hold. We also indicate
the model-independent upper limits on annihilation rates in these
channels as functions of the DM particle mass, obtained from cosmic
ray antiproton data \cite{Cuoco:2017iax}. The regions bounded by the
upper and lower limits represent the area where DM annihilation in
this galaxy can certainly be detected within 100 hours. For Draco, with the NFW profile
and a galactic magnetic field ($B$) of 1.0 $\mu G$, $D_0 = 3 \times
10^{28} \mbox{cm}^2 \mbox{s}^{-1}$ and $\gamma$ = 0.3 \cite{Colafrancesco:2006he}, all of our
benchmark points whose samples are shown as black spots (mostly beyond the LHC reach \cite{LHC_reach, LHC_reach2}) fall in the
detectable range. Remarkably, this pushes the radio search limit up to
$m_{\chi^0_1}$ $\sim$ 8.5 TeV. The reach goes up to even 10 TeV if there
is substantial annihilation in the $b\bar{b}$ channel.

\begin{figure*}[ht!]
\centering
  \includegraphics[height=0.35\textwidth, angle=0]{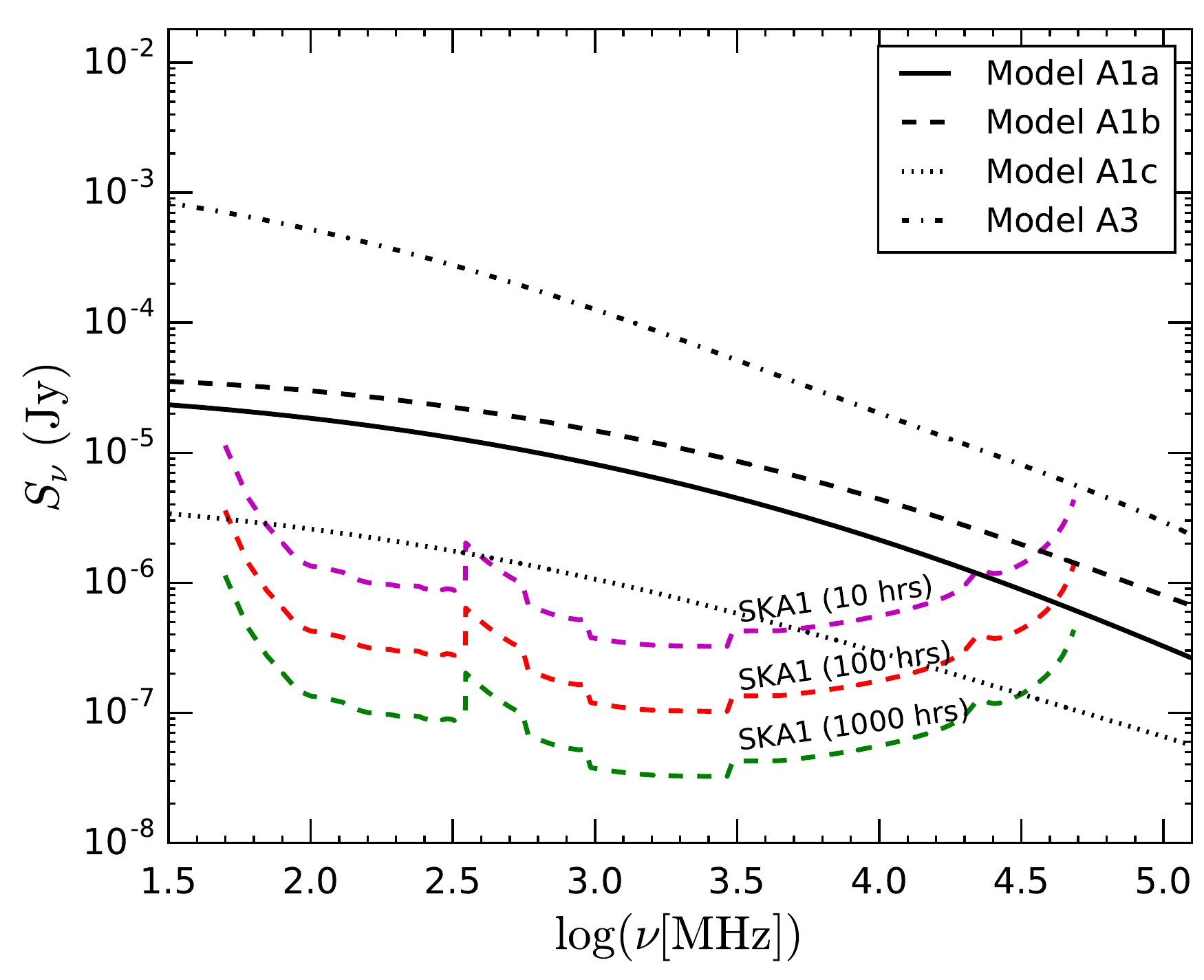}\hspace{8mm}%
  \includegraphics[height=0.35\textwidth, angle=0]{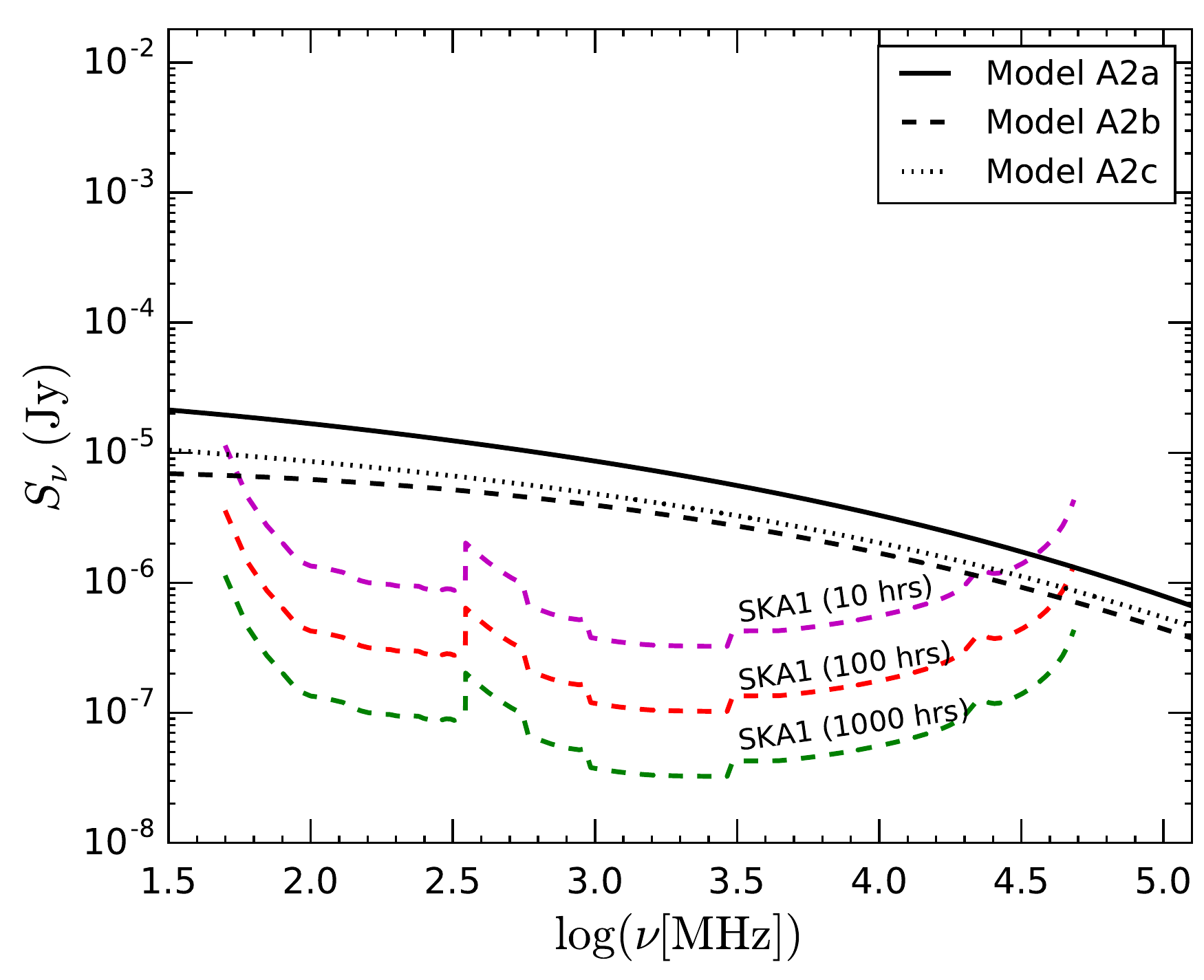}
  \includegraphics[height=0.35\textwidth, angle=0]{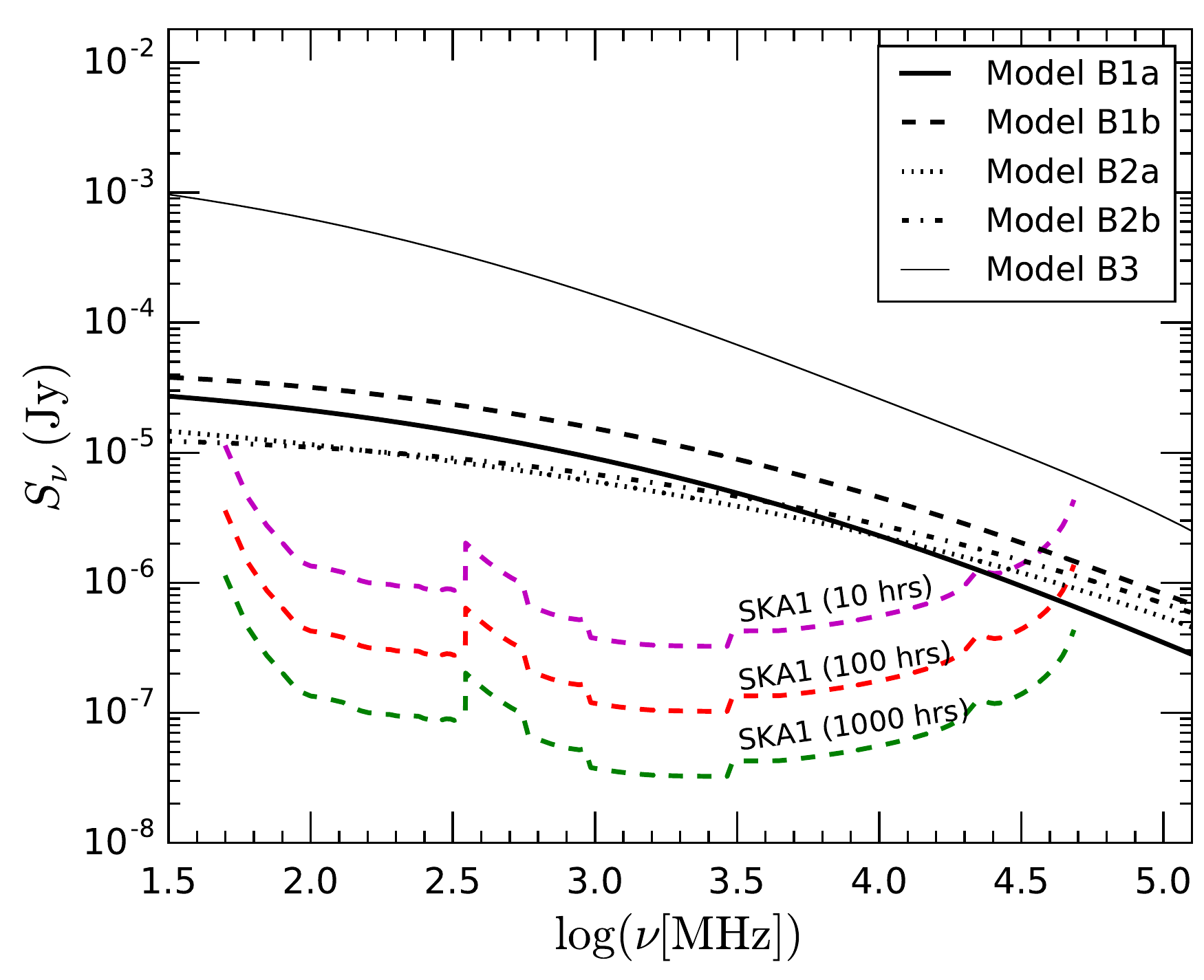}\hspace{8mm}%
  \includegraphics[height=0.35\textwidth, angle=0]{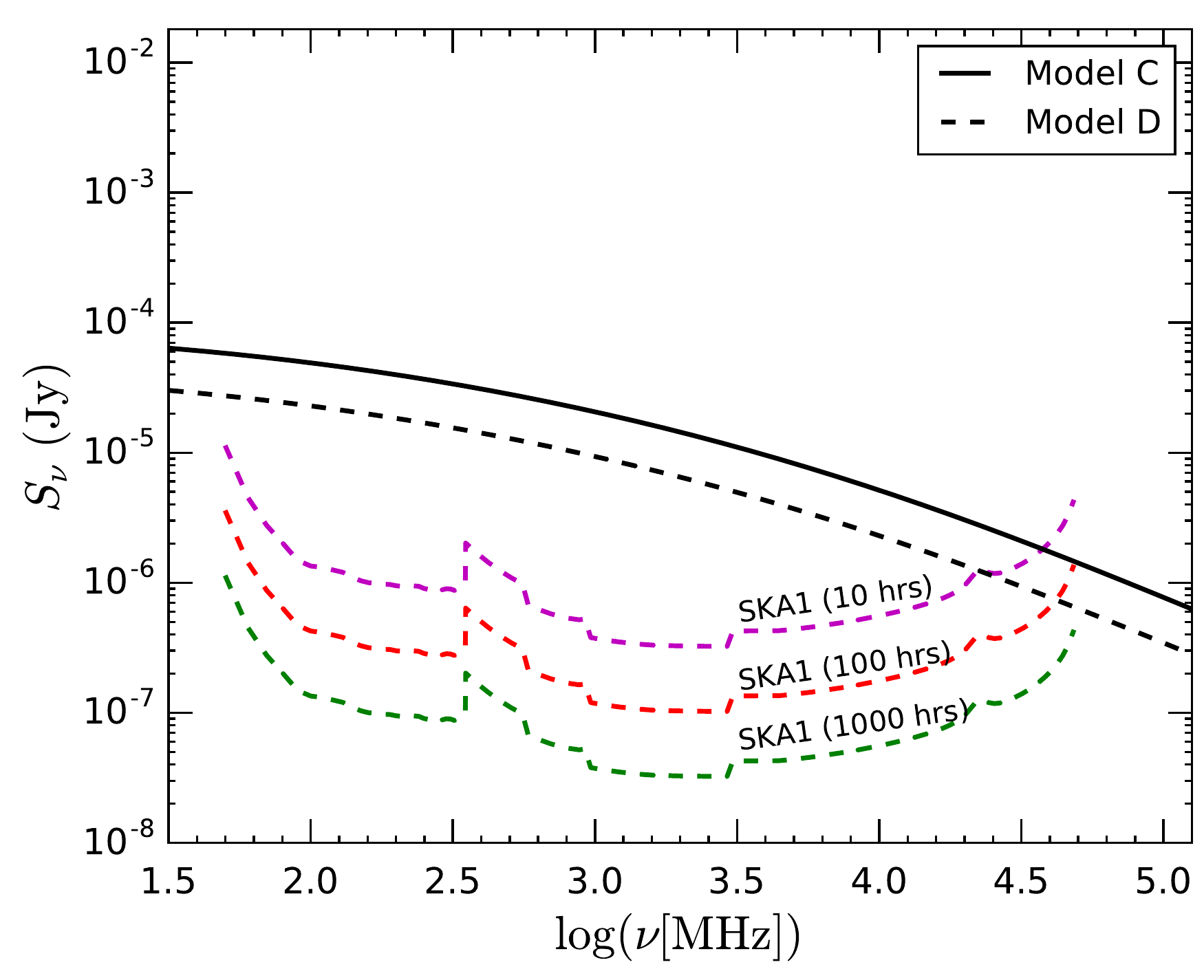}%
  \caption{Synchrotron fluxes for various models (listed in Table
    \ref{Table_benchmark_1} and \ref{Table_benchmark_2}) in the Draco
    dSph galaxy ($D_0 = 3 \times 10^{28} \mbox{cm}^2 \mbox{s}^{-1}$, $B
    = 1$ $\mu G$). The SKA1 sensitivity curves for 10, 100 and 1000 hrs
    are also shown for bandwidth = 300 MHz.}
\label{various_Cases}
\end{figure*}

The frequency spectra of the predicted radio signals
  are shown in Figure \ref{various_Cases}. The expected SKA1
  sensitivities in the frequency range 350 MHz -- 50 GHz
  \cite{SKA} are also shown for observations over 10, 100 and
  1000 hours.\footnote{It is possible that the SKA1
    design may undergo minor changes in the future, leading to gaps in
    the frequency coverage and revisions in the sensitivity
    estimates. This should not affect our main conclusion, since 
    the predicted signals are well above the
    sensitivity limits.} Although the curves are drawn using the NFW
profile for Draco, we have checked that the predictions remain very
similar for other profiles such as Burkert and D05
\cite{Colafrancesco:2006he}. Also, we have assumed no halo
substructures which can in principle enhance the flux even further
\cite{Beck:2015rna}. As per current
understanding, significant radio signals from
from astrophysical processes are unlikely, as dSphs are
mostly devoid of gas and have almost no intrinsic sources of high
energy $e^{\pm}$. The other possible sources of
  contamination are the astrophysical foregrounds, however, they too
  are expected to be sub-dominant for the SKA1 as the large baselines
  will help in resolving out these objects. On the whole, detection is
almost certain for each case within 100 hours; there are several
benchmark points where even 10 hours should suffice. 
Note that the flux depends on $m_{\chi^0_1}$, $\langle
\sigma v \rangle$ and $B_f$. Thus MSSM dynamics crucially decides
detectability.  Overall, the SKA1 clearly goes
beyond the LHC in SUSY-DM search \cite{LHC_reach, LHC_reach2}. As Figure \ref{sv_mchi}
shows, a neutralino DM with mass on the order of 10 TeV (or perhaps
more) may be rendered visible in the process.

\begin{figure*}[ht!]
\centering
  \includegraphics[height=0.35\textwidth, angle=0]{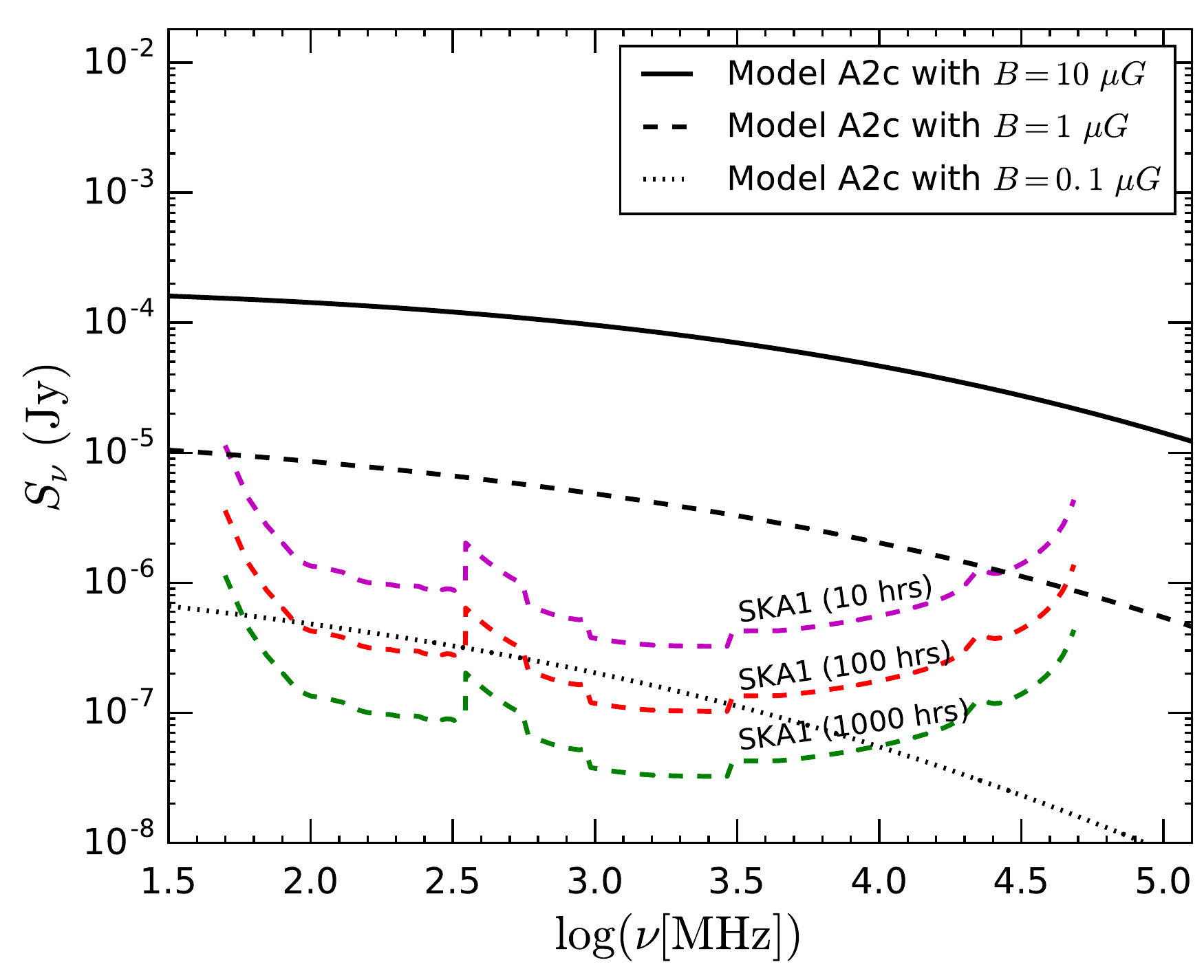}\hspace{8mm}%
  \includegraphics[height=0.35\textwidth, angle=0]{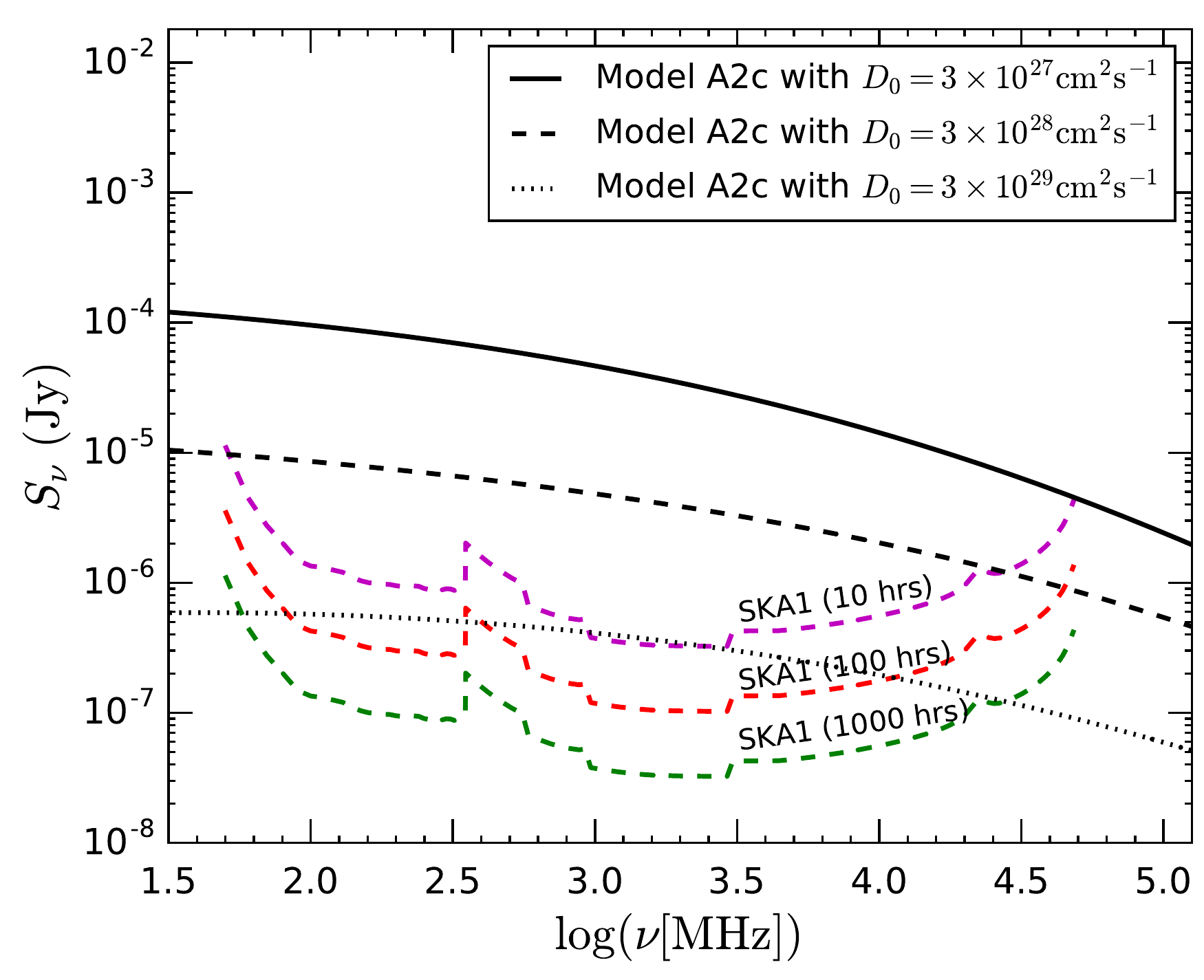}
  \caption{{\it Left:} Synchrotron fluxes for Model A2c for Draco
    ($D_0 = 3 \times 10^{28} \mbox{cm}^2 \mbox{s}^{-1}$)
    with various magnetic fields $B$ = 10.0, 1.0 and 0.1 $\mu G$. {\it
      Right:} Synchrotron fluxes for Model A2c for Draco 
    ($B = 1$ $\mu G$) with different $D_0 =
    3 \times 10^{27}, 3 \times 10^{28}$ and $3 \times 10^{29}
    \mbox{cm}^2 \mbox{s}^{-1}$.}
\label{Case_A2c_B_D0}
\end{figure*}

While the above results are presented for $B = 1.0$ $\mu G$ 
(typical of a dSph like Draco where the magnetic field
has been measured \cite{Chyzy}), the predictions with
other values, namely, $B = 10.0$ and $0.1$ $\mu G$, are  presented
in Figure \ref{Case_A2c_B_D0} (left). We thus see that even for the
pessimistic value of 0.1 $\mu G$, the signals are detectable
up to $10^{3.4} (10^{3.8})$ MHz for 100 (1000) hours of
observation. Figure \ref{Case_A2c_B_D0} (right) shows the effect of
different $D_0$. We once more include the `unfavourable' value of 
$D_0 = 3\times 10^{29} \mbox{cm}^2 \mbox{s}^{-1}$, $\gamma = 0.3$, 
for which detectability should be rather high in the range 
$10^{2-4}$ MHz, for a neutralino
mass $\sim$ 4 TeV, with the coloured particle masses at 10 TeV.

\begin{figure}[ht!]
\begin{center}
\includegraphics[height=0.35\textwidth, angle=0]{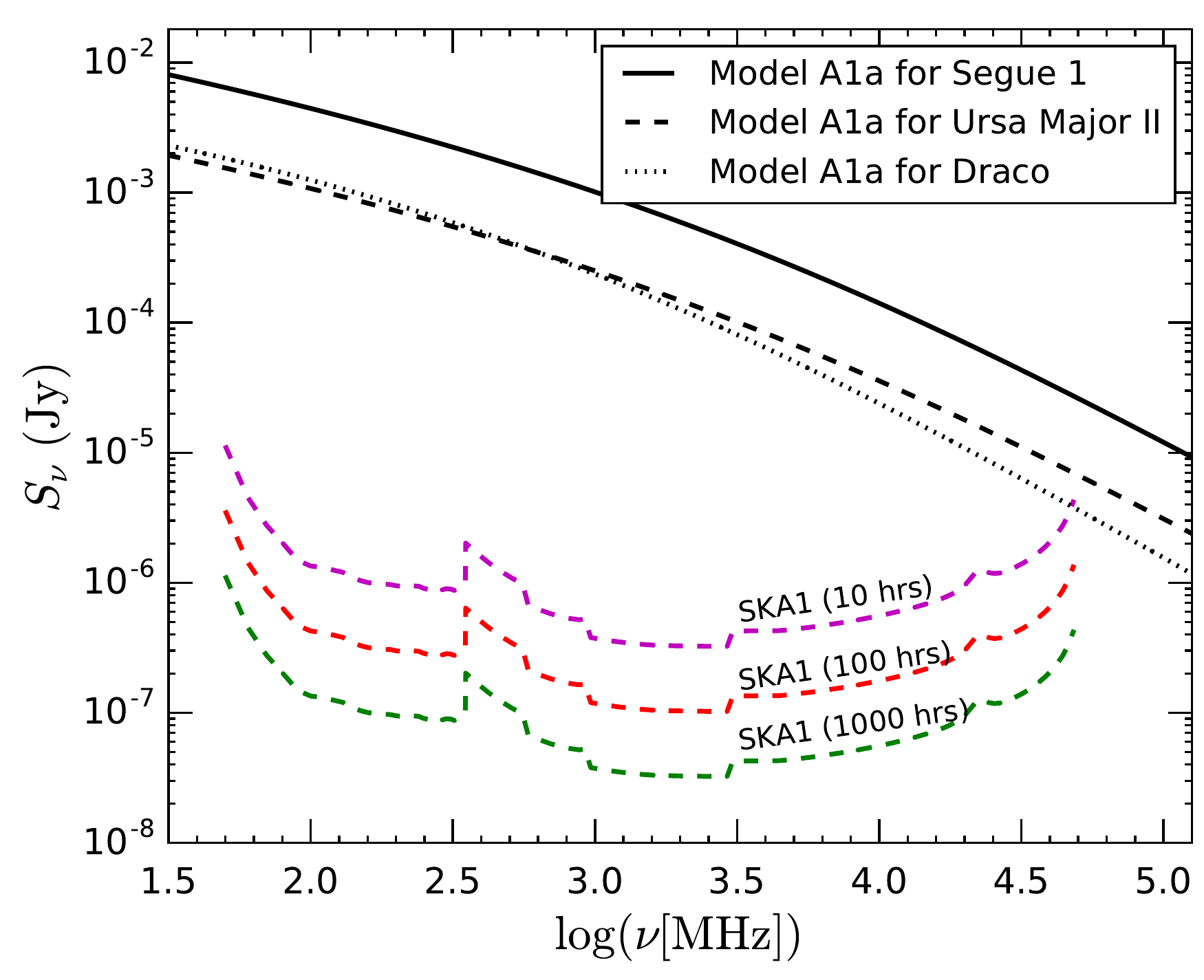}
\caption{Comparison of synchrotron fluxes in Model A1a among 
  Draco, Ursa Major~II and Segue~1 with $D_0 = 3 \times 10^{26} \mbox{cm}^2 \mbox{s}^{-1}$ 
  and  $B = 1$ $\mu G$.}
\label{Case_A1a_galaxy}
\end{center}	
\end{figure}

We finally show in Figure \ref{Case_A1a_galaxy} some predictions for
galaxies nearer than Draco, namely, Ursa Major~II and Segue~1, with $B
= 1.0$ $\mu G$, $D_0 = 3\times 10^{26} \mbox{cm}^2 \mbox{s}^{-1}$, and
appropriate values of $\gamma$ \cite{Colafrancesco:2006he, McDaniel:2017ppt, Natarajan:2013dsa,
  Natarajan:2015hma}. Benchmark A1a is used for illustration. While
detectability is much above threshold here, a comparison 
with Draco tells us that
Segue~1 and Ursa Major~II hold high hopes for DM annihilation
detection, even with larger $m_{\chi^0_1}$. Even if
  SKA1 succeeds in setting upper limits on the flux for
  for most of our benchmark points, it will be possible to
  probe and constrain regions of hitherto unexplored regions in 
  the MSSM parameter space well. Observations of the signal in
  different wavebands, say, radio and $\gamma$-ray frequencies, from
  any dSph may enable also us to break the degeneracies between the MSSM
  parameters and $B, D_0$.

\begin{table}[ht!]
\centering
\begin{tabular}{|c|c|c|c|c|c|c|c|}
\hline 
Model & $\langle \sigma v \rangle$ & BSM particles dominantly \\
      & ($10^{-26} \mbox{cm}^3 \mbox{s}^{-1}$) &  responsible for annihilation, \\
      &                                        &  and their masses(GeV)\\
\hline
A1a & 0.27 & $A^0$(2113.0), $\widetilde{\tau_1}$(5497.0)\\ 
\hline
A1b & 0.77 & $\widetilde{\chi_1^{\pm}}$(1163.5), $\widetilde{\chi_2^0}$(1163.4)\\ 
\hline 
A1c & 0.05 & $\widetilde{\tau_1}$(1037.0), $A^0$(2200.0)\\
\hline 
A2a & 1.76 & $A^0$(7001.2), $\widetilde{\tau_1}$(5490.0)\\  
\hline 
A2b & 1.01 & $\widetilde{\chi_1^{\pm}}$(3085.5), $A^0$(6517.9)\\
\hline
A2c & 1.47 & $A^0$(8293.9), $\widetilde{\chi_1^{\pm}}$(4090.8), \\
    &      & $\widetilde{\tau_1}$(5487.2)\\ 
\hline
A3 & 1.16 & $A^0$(690.8)\\
\hline
B1a & 0.3 & $A^0$(2113.0), $\widetilde{b_1}$(5159.4), \\
    &     & $\widetilde{t_1}$(5047.0), $\widetilde{\tau_1}$(4496.7)\\
\hline
B1b & 0.79 & $\widetilde{\chi_1^{\pm}}$(1153.8), $\widetilde{\chi_2^0}$(1154.8)\\
\hline
B2a & 1.19 & $A^0$(6820.0), $\widetilde{b_1}$(5150.6), \\
    &      & $\widetilde{\tau_1}$(4488.3), $\widetilde{t_1}$(5015.2) \\  
\hline 
B2b & 1.3 & $\widetilde{\chi_1^{\pm}}$(2662.2), $A^0$(5752.4)\\  
\hline
B3 & 1.3 & $A^0$(668.0)\\  
\hline
C & 0.69 & $A^0$(2033.5), $\widetilde{t_1}$(2041.0), \\
  &      & $\widetilde{b_1}$(5085.5), $\widetilde{\tau_1}$(4497.3)\\
\hline
D & 0.31 & $A^0$(2047.0), $\widetilde{t_1}$(1788.0), \\
  &      & $\widetilde{\tau_1}$(5497.5)\\
\hline
E & 9.12 & $A^0$(17035.0)\\
\hline
F & 3.83 & $A^0$(20000.0)\\
\hline
\end{tabular}
\caption{Annihilation rates ($\langle \sigma v \rangle$) for all the benchmark points along with the list of (super)particles dominantly responsible for $\chi^0_1$ DM annihilation for any particular benchmark point.}
\label{Table_benchmark_3}
\end{table}

Table \ref{Table_benchmark_3} shows the annihilation cross-sections of the $\chi^0_1$ 
DM for all our benchmark points, for which the corresponding $\chi^0_1$ masses are supplied in Table \ref{Table_benchmark_2}. Side by side, the new particles apart from $\chi^0_1$, which play the most crucial roles in annihilation are listed, along with their masses for the corresponding benchmarks. As we can see, such role is mostly played among superparticles by the $\chi^{\pm}_1$ and $\chi^0_2$ on the one hand, and the $\widetilde{t_1}$, $\widetilde{b_1}$, and 
$\widetilde{\tau_1}$, on the other. The masses for these particles evidently lie beyond the reach of the LHC for an integrated luminosity of 3000 $\mbox{fb}^{-1}$.

For some benchmark points, the neutral pseudoscalar Higgs ($A^0$) has an important role. While the corresponding $M_A$ in each case is currently allowed by all data, they (with the appropriate values of $tan\beta$ as in Table \ref{Table_benchmark_2}) are unlikely to be seen in any direct signal even at the high-luminosity run.

It is true that some representative MSSM parameters values have included in our sixteen benchmarks. It is of course possible to carry out a similar analysis with a simplified SUSY model containing fewer free parameters. However, we have not restricted ourselves to such scenarios, since our main point is made more emphatically in the general case. The point is that {\em the general MSSM parameter space contains a wide range when not only the $\chi^0_1$ DM but also the (super)particles responsible for its annihilation within dSph's lie beyond the LHC reach.} This is evident form Table \ref{Table_benchmark_3}; one is now looking at scenarios with $m_{\chi^{\pm}_1}$, $m_{\chi^0_2}$ $\in$ [1150 GeV, 4100 GeV], $m_{\widetilde{\tau_1}}$ $\in$ [1000 GeV, 5500 GeV], $m_{\widetilde{t_1}}$ $\in$ [1750 GeV, 5000 GeV], 
$m_{\widetilde{b_1}}$ $\simeq$ 5000 GeV, $M_A$ $\in$ [670 GeV, 20000 GeV]. All of these (excepting perhaps $M_A$ $\simeq$ 670 GeV) are beyond the LHC reach \cite{LHC_reach, LHC_reach2, Aaboud:2017sjh}.

We thus conclude that the SKA1, mostly
with 100 hours of observation, should be able to detect radio synchrotron
signals of MSSM DM annihilation, for cases where the superparticle
masses are well above the reach of the LHC. Even neutralinoes below a TeV, which the LHC cannot probe due to overwhelming backgrounds, are covered by such observation. This holds
even for conservative values of astrophysical parameters, and thus
underscores a new potential of the SKA.\\

\noindent
{\bf Acknowledgements:}\\
 We thank AseshKrishna Dutta and Subhadeep Mondal 
 for helpful comments. AK and BM were partially supported
 by funding available from the Department of Atomic Energy, Government
 of India, for the Regional Centre for Accelerator-based Particle
 Physics (RECAPP), Harish-Chandra Research Institute. TRC and SM acknowledge
 the hospitality of RECAPP.




\end{document}